\documentclass[aps,twocolumn,showpacs,preprintnumbers,prl,superscriptaddress,10pt,nofootinbib]{revtex4-1}
\usepackage[utf8]{inputenc}
\usepackage{graphicx}
\usepackage{amsmath,amssymb}
\usepackage{amsfonts}
\usepackage{physics}
\usepackage{xspace}
\usepackage{bm}
\usepackage[normalem]{ulem}
\usepackage{mathrsfs}
\usepackage[dvipsnames]{xcolor}
\usepackage[unicode]{hyperref}
\hypersetup{colorlinks=true, citecolor=MidnightBlue,
	linkcolor=Brown, urlcolor=Blue, linktocpage=true}
\usepackage[all]{hypcap}
\usepackage{multirow}
\usepackage[format=plain,justification=RaggedRight,labelsep=endash,font=small,labelfont=sc]{caption}
\usepackage[export]{adjustbox}

\usepackage{caption}
\usepackage{subcaption}

\definecolor {darkgreen}{rgb}{0.2,0.7,0.2}

\definecolor {dark}{rgb}{0.43,0.5,0.5}

\newcommand{\eq}{\begin{equation}}
\newcommand{\be}{\begin{equation}}
\newcommand{\eeq}{\end{equation}}
\newcommand{\ee}{\end{equation}}
\newcommand{\de}{\partial}

\newcommand{\trK}{\mathrm{tr}K}
\newcommand{\trT}{\left(T^{\phi}+T\right)}
\newcommand{\trKhat}{\mathrm{tr}{\hat K}}

\newcommand{\trAtilde}{\mathrm{tr}{\bar A}}

\begin{document}

\title{No evidence of kinetic screening in simulations of merging binary neutron stars beyond general relativity}
 \author{Miguel Bezares}
 \affiliation{SISSA, Via Bonomea 265, 34136 Trieste, Italy and INFN Sezione di Trieste}
 \affiliation{IFPU - Institute for Fundamental Physics of the Universe, Via Beirut 2, 34014 Trieste, Italy}
 \author{Ricard Aguilera-Miret}
 \affiliation{Departament  de  F\'{\i}sica,  Universitat  de  les  Illes  Balears  and  Institut  d'Estudis Espacials  de  Catalunya,  Palma  de  Mallorca,  Baleares  E-07122,  Spain}
\affiliation{Institut Aplicacions Computationals (IAC3),  Universitat  de  les  Illes  Balears,  Palma  de  Mallorca,  Baleares  E-07122,  Spain}
  \author{Lotte ter Haar}
 \affiliation{SISSA, Via Bonomea 265, 34136 Trieste, Italy and INFN Sezione di Trieste}
 \affiliation{IFPU - Institute for Fundamental Physics of the Universe, Via Beirut 2, 34014 Trieste, Italy}
 \author{Marco Crisostomi}
 \affiliation{SISSA, Via Bonomea 265, 34136 Trieste, Italy and INFN Sezione di Trieste}
 \affiliation{IFPU - Institute for Fundamental Physics of the Universe, Via Beirut 2, 34014 Trieste, Italy}
  \author{Carlos Palenzuela}
\affiliation{Departament  de  F\'{\i}sica,  Universitat  de  les  Illes  Balears  and  Institut  d'Estudis Espacials  de  Catalunya,  Palma  de  Mallorca,  Baleares  E-07122,  Spain}
\affiliation{Institut Aplicacions Computationals (IAC3),  Universitat  de  les  Illes  Balears,  Palma  de  Mallorca,  Baleares  E-07122,  Spain}
  \author{Enrico Barausse}
 \affiliation{SISSA, Via Bonomea 265, 34136 Trieste, Italy and INFN Sezione di Trieste}
 \affiliation{IFPU - Institute for Fundamental Physics of the Universe, Via Beirut 2, 34014 Trieste, Italy}

\begin{abstract}
We have conducted 
fully relativistic simulations 
in a class of scalar-tensor theories with derivative self-interactions
and screening of local scales.
By using high-resolution shock-capturing
methods and a non-vanishing shift vector, we
have managed to avoid issues plaguing similar attempts in the past.
We have first confirmed recent results by ourselves in spherical symmetry, obtained with an approximate approach and pointing at a partial breakdown of the screening in black-hole collapse.
Then, we  considered 
the late inspiral and merger of 
binary neutron stars. We found that screening tends to suppress the (subdominant) dipole scalar emission, but 
not the (dominant) quadrupole scalar mode.
Our results point at quadrupole
scalar signals as large as (or even larger than) in Fierz-Jordan-Brans-Dicke theories with the same conformal coupling, for strong-coupling scales in the MeV range that we can simulate.
\end{abstract}
\date{\today \hspace{0.2truecm}}

\maketitle
\flushbottom

The investigation of gravitational theories beyond General Relativity (GR) has recently intensified, boosted by the detection of gravitational waves (GWs) by LIGO/Virgo~\cite{TheLIGOScientific:2016src,Abbott:2018lct,LIGOScientific:2019fpa,Abbott:2020jks}, which allows 
for testing gravity in
the hitherto unexplored strong-gravity and highly relativistic regime. Natural questions 
preliminary to these tests, however, are the following:
Do we really need to modify GR? What are the open problems that GR cannot address and which we wish to (at least partially) solve with a different theory? 
Leaving aside  quantum gravity completions of GR, whose
modifications only become important at the Planck scale, 
GR cannot explain the late-time accelerated expansion of the Universe,
unless one introduces a cosmological constant (with its associated problems) or a dark energy component.
Therefore, an alternative gravity theory should provide $\sim {\cal O}(1)$ effects on cosmological scales to improve upon GR. 

However, $\sim {\cal O}(1)$ effects on large (cosmological) scales typically imply also $\sim {\cal O}(1)$ deviations from GR on local  (e.g. solar-system~\cite{Will:2014kxa} and binary-pulsar~\cite{Damour:1991rd,Kramer:2006nb,Freire:2012mg}) scales, where  GR is tested to within $\lesssim {\cal O}(10^{-5})$. To comply with these stringent constraints, an obvious possibility is
that the additional gravitational polarizations (besides the tensor modes of GR) have sufficiently weak self-couplings and coupling with other fields, including matter. This is the case of e.g. Fierz-Jordan-Brans-Dicke (FJBD) theory~\cite{Fierz:1956zz,Jordan:1959eg,Brans:1961sx}. In this way, however,  their 
cosmological effects are lost, and the theory cannot provide an effective dark-energy phenomenology.
Less trivially, the non-tensor gravitons may self-interact (via derivative operators) so strongly near matter sources that the ``fifth'' forces they mediate are locally suppressed. This is the idea behind kinetic (or $k$-mouflage) \cite{Babichev:2009ee} and Vainshtein \cite{Vainshtein:1972sx} screening.

Although screening mechanisms have been widely advocated to reconcile  modifications of GR on cosmological scales with local tests (see e.g.~\cite{Clifton:2011jh,Koyama:2015vza} for reviews), their validity has never been proven beyond certain simplified approximations (single body \cite{Babichev:2009ee}, weak gravity \cite{Babichev:2010jd,Babichev:2013pfa}, quasistatic configurations \cite{deRham:2012fg,deRham:2012fw,Dar:2018dra}, spherical symmetry \cite{Crisostomi:2017lbg}), let alone in realistic compact-binary coalescences. Here, we will provide the first numerical-relativity simulations of compact binaries in theories with kinetic screening, and finally address the long-standing question of whether screening mechanisms 
render GW generation
 indistinguishable from GR.

Theories with kinetic screening, i.e. $k$-essence theories, are the most compelling modified-gravity candidate for dark energy after GW170817 \cite{Monitor:2017mdv,TheLIGOScientific:2017qsa} and other constraints related to GW propagation \cite{Creminelli:2018xsv, Creminelli:2019kjy, Babichev:2020tct}. In the Einstein frame~\cite{PhysRevD.1.3209}, the action
of this scalar-tensor theory is~\cite{Chiba:1999ka, ArmendarizPicon:2000dh}
\be
S=\int \mathrm{d}^{4}x\sqrt{-\tilde{g}}\left[\frac{M_{\mathrm{Pl}}^{2}}{2}\tilde{R} + K(\tilde{X})  \right]  + S_{m}\left[\frac{\tilde{g}_{\mu\nu}}{\Phi(\phi)},\Psi_m\right] \,, \label{action}
\ee
where $\tilde{X}\equiv \tilde{g}^{\mu\nu} \partial_\mu\phi \partial_\nu\phi$, $\Psi_m$ are the matter fields,
$\Phi=\exp\left(\sqrt{2}\,\alpha\, \phi/M_{\rm Pl}\right)$, being $M_\mathrm{Pl}=(8\pi G)^{-1/2}$ the Planck mass, and we set $\hbar=c=1$.
For $K(\tilde{X})$, we only consider  the lowest-order terms
\be
K(\tilde{X})=-\frac{1}{2}\tilde{X}+\frac{\beta}{4\Lambda^4}\tilde{X}^2-\frac{\gamma}{8\Lambda^8}\tilde{X}^3 + \dots \,,\label{kessence}
\ee
where $\Lambda$ is the strong-coupling scale of the effective field theory (EFT).
For the 
scalar to be responsible for dark energy, one needs $\Lambda_\mathrm{DE}\sim(H_0 M_\mathrm{Pl})^{1/2}\sim 2 \times 10^{-3}\;\mathrm{eV}$, where $H_0$ is the present Hubble rate.
The conformal coupling $\alpha$  and the coefficients $\beta$ and $\gamma$ appearing in Eq.~(\ref{kessence}) are dimensionless and $\sim\mathcal{O}(1)$.

Performing neutron-star (NS) numerical-relativity simulations in $k$-essence is complicated by the strong-coupling nature of the scalar self-interactions\footnote{For binary black holes, instead, no deviations from GR 
are expected, since the scalar  equation's  source ($\propto \alpha T$) vanishes. This is also the reason behind the no-hair theorem for isolated black holes in $k$-essence~\cite{Hui:2012qt}.  The only deviations from GR in vacuum and in absence of a scalar mass~\cite{Detweiler:1980uk}
may occur because of non-trivial initial conditions (which lead to transient scalar effects~\cite{Healy:2011ef}) and time-dependent boundary conditions for $\phi$~\cite{Jacobson:1999vr}.}. Although the  Cauchy problem  is  locally  well-posed, the evolution equations may change character from hyperbolic to parabolic and then elliptic at finite time \cite{Bernard:2019fjb,Bezares:2020wkn}. This behavior can  be avoided by imposing ``stability'' conditions on the coefficients $\beta, \gamma, \ldots$  of Eq.~(\ref{kessence}) \cite{Bezares:2020wkn}, just like other conditions (e.g. no ghosts, tachyons or gradient instabilities) are usually imposed on the coefficients of any EFT\footnote{The radiative stability of these conditions in the non-linear regime has been shown in \cite{deRham:2014wfa,Brax:2016jjt}.}. Nevertheless, even under these stability conditions,
the characteristic speeds of the scalar field  are found to  diverge  during  gravitational  collapse \cite{Bernard:2019fjb,Figueras:2020dzx,Bezares:2020wkn,terHaar:2020xxb,Bezares:2021yek}. This 
may be physically pathological, and it is certainly a serious practical drawback, as 
 the Courant–Friedrichs–Lewy (CFL) condition~\cite{1928MatAn.100...32C} forbids 
to evolve the fully non-linear dynamics past this divergence. Recently,
 \cite{Bezares:2021yek} 
 managed to evolve gravitational collapse past the scalar-speed divergence 
 by slightly modifying the dynamics, with the addition of an extra driver field~\cite{Cayuso:2017iqc,Allwright:2018rut,Cayuso:2020lca}. This technique, while  {\it ad hoc} and approximate,
suggests that the divergence is not of physical origin, but rather linked to the gauge choice, as conjectured in~\cite{Bezares:2020wkn}. In this work, we  use indeed
a gauge choice (including a non-vanishing shift) that
 maintains the 
characteristic speeds finite during both gravitational collapse and binary evolutions
in $3+1$ dimensions.
The latter constitute the first fully dynamical simulations of the GW generation by binary systems in theories with screening.

\begin{figure}
    \includegraphics[width=0.45\textwidth]{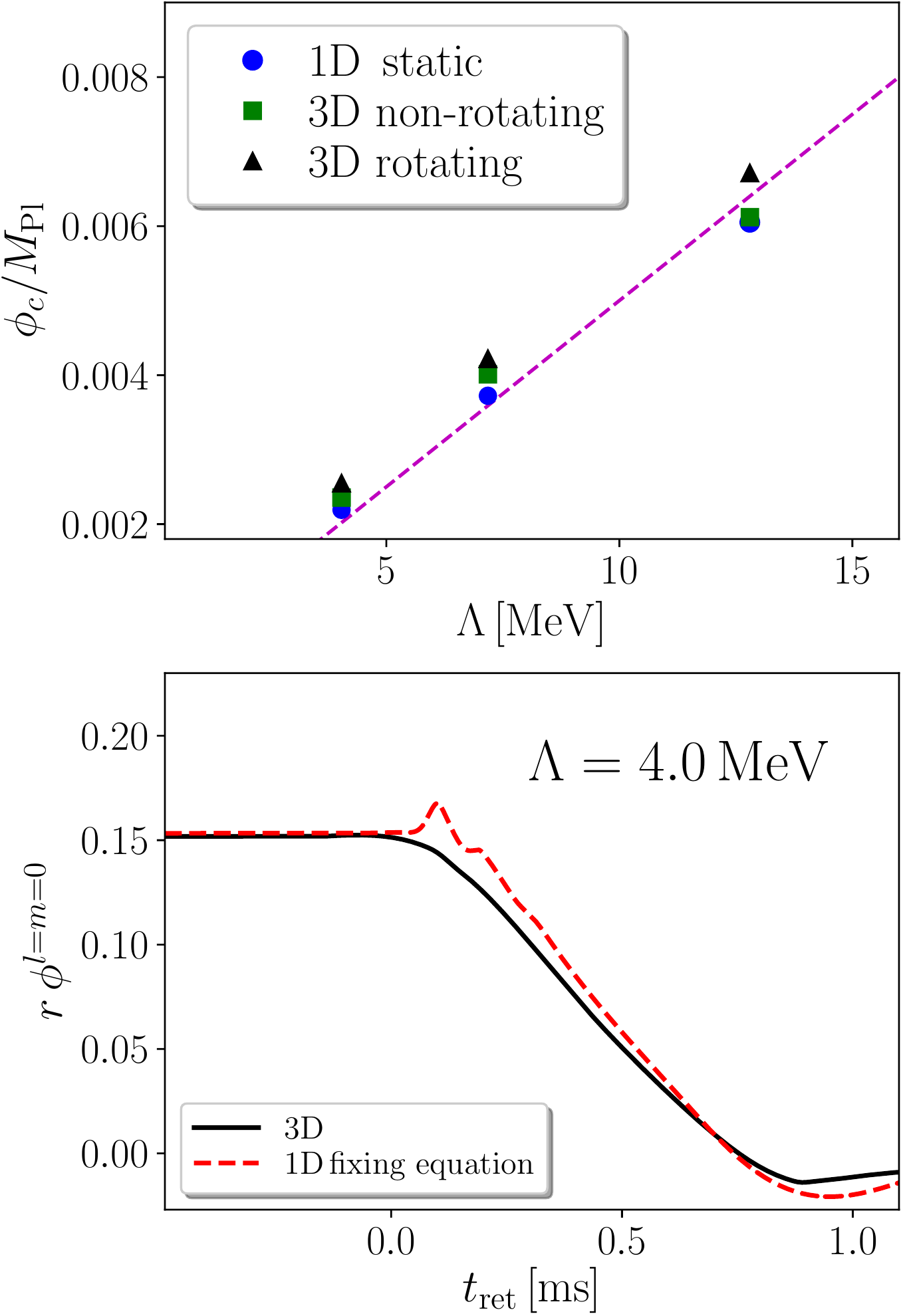}	
	\caption{{\it Top}: Central scalar field for rotating/non-rotating stars produced with our three-dimensional code and  non-rotating ones produced with the static one-dimensional code of~\cite{terHaar:2020xxb}. Stellar masses are $M\approx 1.74\,M_{\odot}$. Also shown is the expected linear scaling with $\Lambda$~\cite{Bezares:2021yek}. {\it Bottom}:  Scalar field at the extraction radius for gravitational collapse (with mass $M\approx 1.74\,M_{\odot}$), obtained with the one-dimensional code of~\cite{Bezares:2021yek} (using an approximate fixing-equation approach) and our three-dimensional code.}
	\label{fig:1d}
\end{figure}

\textit{Set-up:} In order to perform fully relativistic numerical simulations of binary mergers in $k$-essence theory, we consider the Einstein frame and evolve the CCZ4 formulation~\cite{Alic:2011gg,Palenzuela:2018sly,Bezares:2017mzk} of the Einstein equations (with the 1+log slicing~\cite{Bona:1994dr} and the Gamma-driver shift condition~\cite{Alcubierre:2002kk}) coupled to a perfect fluid (adopting an ideal-gas equation of state with $\Gamma=2$)~\cite{Liebling:2020jlq} and a scalar field~\cite{Bezares:2018qwa}. 
Without loss of generality, we fix $\beta=0$ and $\gamma=1$, which ensure the well-posedness of the Cauchy problem~\cite{Bezares:2020wkn} and the existence of screening solutions~\cite{terHaar:2020xxb,Bezares:2021yek}\footnote{This is also the simplest choice that satisfies recent positivity bounds \cite{Davis:2021oce} for a healthy (although unknown) UV completion of the theory.  Those bounds dictate that the leading term in $K(X)$ should have odd power and a negative coefficient. See however \cite{Aoki:2021ffc} for a different claim.}. Furthermore, we set the conformal coupling  to $\alpha\approx0.14$. As discussed in \cite{Bezares:2021yek}, 
 $\Lambda\sim\Lambda_\mathrm{DE}$
is intractable numerically due to the hierarchy
between binary and cosmological scales (which
leads to  $\Lambda_\mathrm{DE}\sim10^{-12}$ in units
adapted to the binary system). Like in \cite{terHaar:2020xxb,Bezares:2021yek}, we study $\Lambda\gtrsim 1\, \mathrm{MeV}$ (for which screening is already present).

\begin{figure*}
    \includegraphics[width=0.99\textwidth]{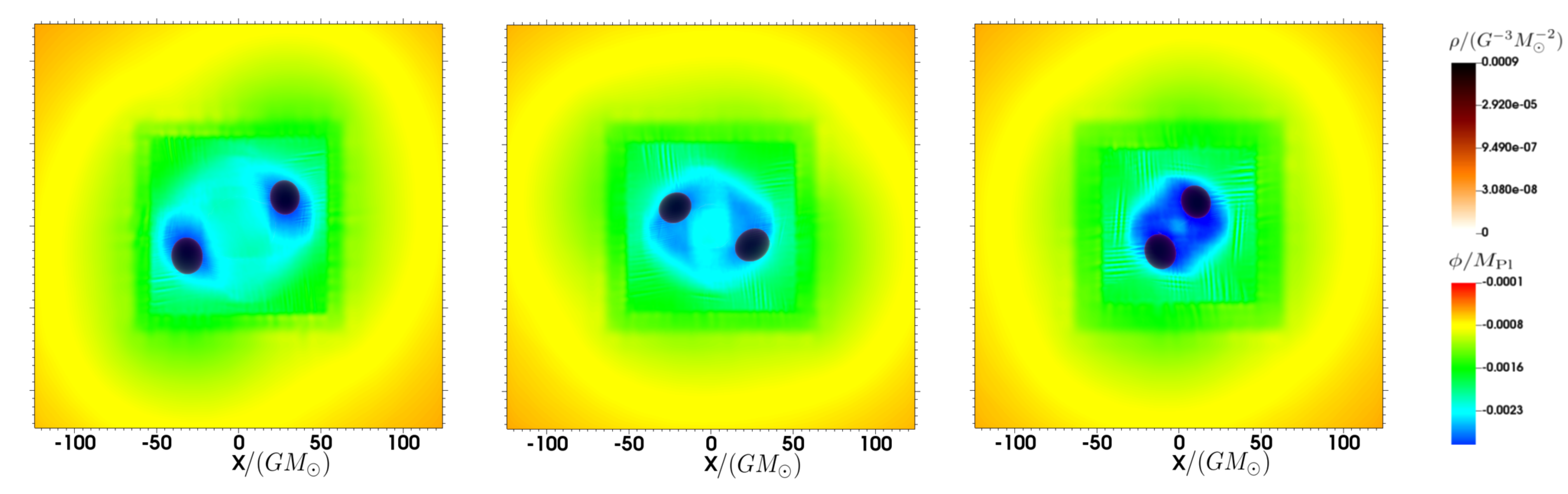}
	\caption{Unequal-mass  ($q=0.91$) NS binary  with $\Lambda=4.0\,\rm{MeV}$, shown at successive times. The color code represents the scalar field, which is initially centered on each star and then develops into a common ``envelope'', while the dark orange represents the fluid density.}
	\label{fig:slice_2D}
\end{figure*}

The computational code, generated by using the platform \textit{Simflowny}~\cite{ARBONA20132321,ARBONA2018170,simflownywebpage}, runs under the SAMRAI infrastructure~\cite{Hornung:2002,Gunney:2016,samraiwebpage}, which provides parallelization and the adaptive mesh refinement (AMR) required to solve the different scales in the problem. 
We use fourth-order finite difference operators to discretize our equations~\cite{Palenzuela:2018sly}. For the fluid and the scalar, we use High-Resolution Shock-Capturing (HRSC) methods to  deal with  shocks, as discussed in~\cite{Bezares:2020wkn,terHaar:2020xxb,Bezares:2021yek}. A similar GR code, using the same methods, has been recently used to simulate binary NSs~\cite{Liebling:2020jlq,Liebling_2021}.
Our computational domain ranges from $[-1500,1500]^3\,\rm{km}$ and contains 6 refinement levels. Each level has twice the resolution of the previous one, 
achieving a resolution of $\Delta x_{6}= 300\,\rm{m}$ on the finest grid. We use a Courant factor $\lambda_{c}\equiv\Delta t_{l} / \Delta x_{l} = 0.4$ on each refinement level $l$ to ensure stability of the numerical scheme.

\textit{Isolated stars and gravitational collapse:} We construct initial data for NS systems in $k$-essence theory by relaxation~\cite{Barausse:2012da,Palenzuela:2013hsa}, i.e. we generate GR solutions using LORENE~\cite{lorene} and evolve them in $k$-essence  until that they relax to stationary solutions.
These solutions agree with the non-rotating solutions for $k$-mouflage stars found in \cite{terHaar:2020xxb,Bezares:2021yek}, thus validating
our relaxation technique (Fig.\,\ref{fig:1d}, top panel). 
Similarly, we have also considered rotating solutions in $k$-essence,
finding for the first time that {\it (i)} they behave qualitatively like the  non-spinning ones, and {\it (ii)} the screening mechanism also survives in axisymmetry (Fig.\,\ref{fig:1d}, top panel).
We also reproduced the dynamics of stellar oscillations in $k$-essence found in~\cite{terHaar:2020xxb,Bezares:2021yek}, and managed to follow 
the (spherical) black-hole collapse of a NS (Fig.\,\ref{fig:1d}, bottom panel).
These simulations were inaccessible, due to the diverging characteristic speeds, within the framework of~\cite{terHaar:2020xxb},
without the addition of an extra driver field and a ``fixing equation''~\cite{Cayuso:2017iqc,Allwright:2018rut} for it~\cite{Bezares:2021yek}.
Here, using the gauge conditions mentioned above (and typically employed in numerical-relativity simulations of compact objects in 3+1 dimensions), we found no divergence of the characteristic speeds in our class of $k$-essence theories.
The collapse obtained with this gauge matches exactly the results obtained in~\cite{Bezares:2021yek}, as shown in the bottom panel of Fig.\,\ref{fig:1d},  corroborating the (approximate) ``fixing equation'' technique employed there. We will therefore use the aforementioned gauge conditions throughout this paper.

 \textit{Binary evolutions:} Like in the isolated case, initial data for binary systems are constructed by relaxation. The relaxation process occurs approximately in the initial $ 4\,\rm{ms}$ of our simulations, and does not impact significantly the subsequent binary evolution. We consider binary NSs in quasi-circular orbits with a total gravitational mass $2.8-2.9\,M_{\odot}$ and mass ratio $q=M_2/M_1 =[0.72,1]$. Time snapshots of a binary  with mass ratio $q=0.9$ in a theory with $\Lambda\approx 4$ MeV are shown in Fig~\ref{fig:slice_2D}, displaying both the star's density and the scalar field. The screening radii of the stars in isolation are $\sim 120$ km and thus larger than the initial system separation. This is the physically relevant situation, as for $\Lambda\sim\Lambda_{\rm DE}$ the screening radii are $\sim 10^{11}$ km. Also observe the formation of a scalar wake trailing  each star (for the most part), with the two wakes merging in the last stages of the inspiral. 

The response of a detector to the GW signal from NS binaries 
is encoded in the Newman-Penrose invariants in the Jordan frame~\cite{PhysRevD.8.3308}, i.e. projections
of the Riemann tensor on a null tetrad $l,n,m,\bar{m}$
adapted to outgoing waves.
Tensor and scalar GWs are encoded respectively in $\psi_4 =- R_{\ell \bar{m}\ell \bar{m}}=\Phi\, \tilde{\psi}_4$
and in $\phi_{22}=- R_{l {m}l \bar{m}}=\phi\, \left(\tilde{\phi}_{22}- l^\nu l^\mu \nabla_{\nu} \nabla_{\mu} \log \Phi/2 + ...\right)$, with a tilde denoting quantities in the Einstein frame (where we perform our simulation)
and the dots denoting  terms subleading in the distance $r$. 
We place our extraction radius outside the screening radius of the 
individual NSs, at distances of 300 km from the center of mass. This is justified because
the distance to the detector is typically $\gg 10^{11}$ km, which
is the screening radius for $\Lambda\sim \Lambda_{\rm DE}$, even for Galactic sources. In this regime, the amplitude of scalar perturbations decays as $1/r$, and 
therefore $\phi_{22} \approx - \alpha \sqrt{16\pi\,G}\partial_t^2\phi + O({1}/{r^2})
$~\cite{Barausse:2012da,Bezares:2021yek}.

\begin{figure*}
    \includegraphics[width=0.96\textwidth]{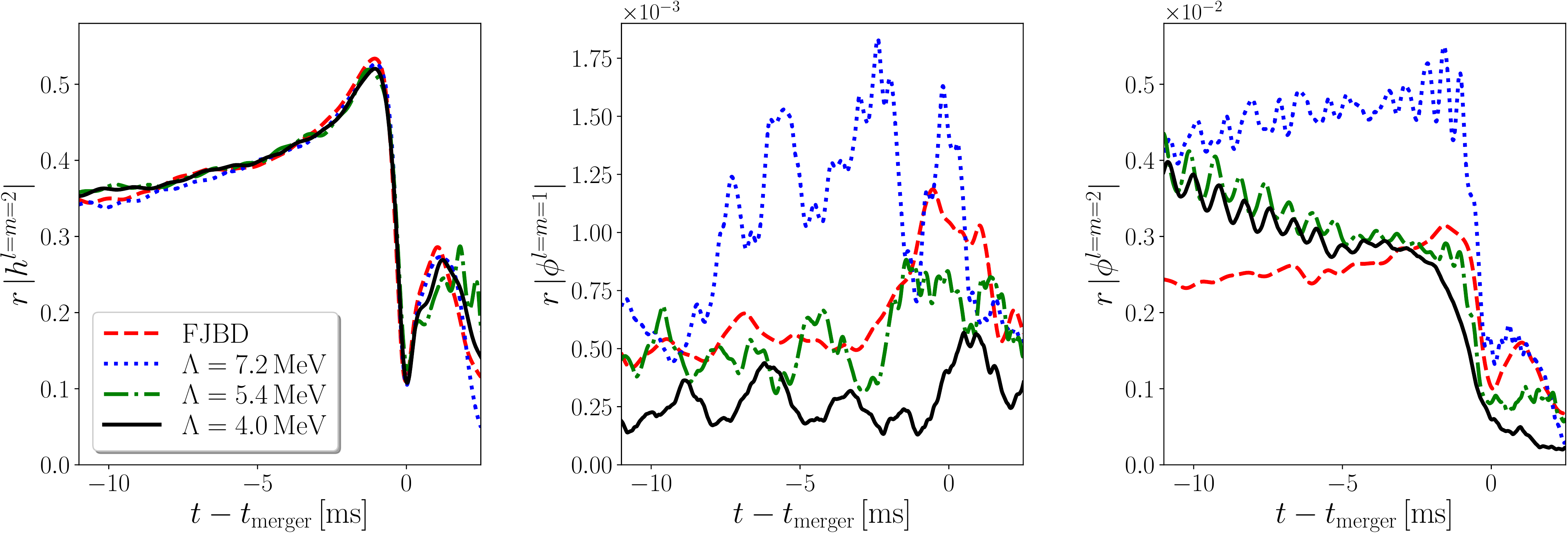}
	\caption{Tensor ($l=m=2$) and scalar ($l=m=1$ and $l=m=2$) strain  for a NS merger with $q=0.91$, in $k$-essence and FJBD.
	\label{fig:amplitude_scalarmodes}}
\end{figure*}

\begin{figure*}
		\includegraphics[width=0.96\textwidth]{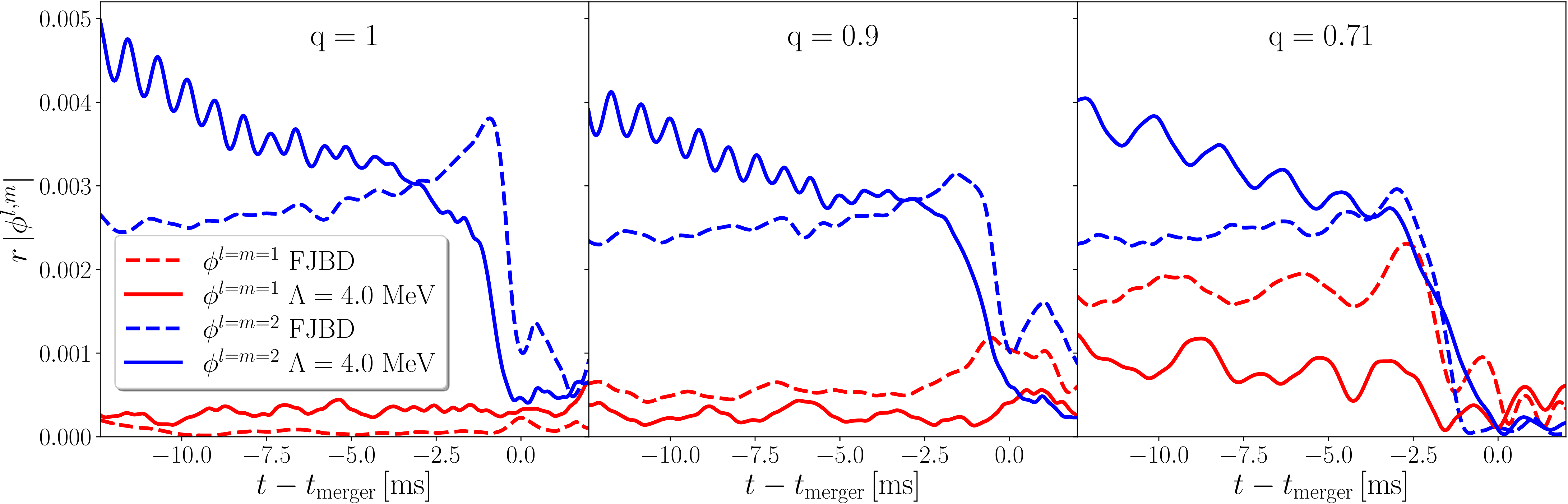}
	\caption{
Dipole ($l=m=1$) and quadrupole ($l=m=2$) scalar strain for merging NS binaries of varying mass ratio, in $k$-essence and FJBD.}
	\label{fig:scalar_radiation}
\end{figure*}

The tensor and scalar ``strains'', $h$ and $h_s$, are defined
by integrating $\psi_4$ and $\phi_{22}$ twice in time, i.e. $\psi_4=\partial_t^2 h/2$ and $\phi_{22}=\partial_t^2 h_s$. The latter definition yields simply $h_s\propto \phi$~\cite{Gerosa:2016fri}. We  then
decompose $h$ and $h_s$ (or $\phi$) into spin-weighted spherical harmonics. As expected, the dominant contribution comes from the $\ell=m=2$ mode for the tensor strain. For 
the scalar emission the monopole $\ell=m=0$ is suppressed,
and the main contribution comes from the dipole ($\ell=m=1$) mode and (mostly) the quadrupole ($\ell=m=2$) mode. The results for four simulations
-- for $\Lambda\approx4$, 5 and 7 MeV and for FJBD (corresponding to $\Lambda\to\infty$), with $\alpha\approx0.14$ -- are shown in Fig.~\ref{fig:amplitude_scalarmodes}.
Notice that we do not show the GR tensor strain as it is practically indistinguishable from the FJBD one on this timescale \cite{Barausse:2012da}.
The three values of $\Lambda$ predict screening radii larger than the initial separation between the stars.

As can be seen, the tensor strains are very similar, even after the merger (corresponding to the peak amplitude). As for the scalar, the suppression
of the $\ell=m=0$ mode is expected, since monopole emission
vanishes in FJBD theory for quasi-circular binaries~\cite{Damour:1992we,Will:1989sk}. The $\ell=m=1$ dipole mode is instead small but non-vanishing, as expected for unequal-mass binaries in FJBD, with signs of screening suppression as $\Lambda$ decreases.
However, the (dominant) $\ell=m=2$ scalar quadrupole mode
is always {\it larger} than in FJBD theory, suggesting that the screening is not effective at suppressing the quadrupole scalar emission in the late inspiral/merger. The amplitude also seems to increase when going to low frequencies/early times, in the simulations with $\Lambda\approx4$ and 5 MeV. Note that
one does not expect a continuous limit to FJBD ($\Lambda\to\infty$) when $\Lambda$ increases. In FJBD there is no screening and the binary is always in the perturbative regime, while in $k$-essence 
the separation is always smaller than the screening radii. This is true even for observed binary pulsars, which have separations $\lesssim 10^5$ km vs screening radii of $\sim 10^{11}$ km for $\Lambda\sim\Lambda_{\rm DE}$.

The dependence on the mass ratio of the binary (which we
set to $q=1$, 0.9 and 0.71) is shown in Fig.~\ref{fig:scalar_radiation}, for the FJBD and $\Lambda\approx 4$ MeV cases. As can be observed, 
quadrupole fluxes are largely unaffected by $q$ in both theories,
with the $k$-essence ones consistently larger, especially at early times. The dipole fluxes in $k$-essence show again signs of
suppression relative to FJBD, at least for $q\neq1$, but in both theories they grow 
as $q$ decreases.  This is expected, since PN calculations in FJBD~\cite{Damour:1992we,Will:1989sk} predict that the dipole amplitude should scale 
as the difference of the stellar scalar charges,
which grows as $q$ decreases. 
For $q=1$, instead, the dipole flux in FJBD is compatible with zero (as predicted by PN theory~\cite{Damour:1992we,Will:1989sk}), while
it does {\it not} vanish in $k$-essence.

\textit{Conclusions:} We have performed for the first time
 fully relativistic simulations of binary NSs in theories of gravity with kinetic screening of local scales. 
 We dealt  with shocks in the scalar field
 by a HRSC method and by  adopting 
 a gauge with non-zero shift 
  that  prevents  divergences of the characteristic speeds, which  plagued previous attempts~\cite{Bernard:2019fjb,Bezares:2020wkn,terHaar:2020xxb}. With this setup, we have confirmed previous preliminary results by ourselves~\cite{Bezares:2021yek}, which were obtained with an approximate fixing-equation approach~\cite{Cayuso:2017iqc,Allwright:2018rut} and which hinted at a possible breakdown of the screening in black-hole collapse. 
  In the late inspiral and merger of binary NSs, 
 the (subdominant) dipole scalar emission is screened (at least for unequal masses), but the (dominant)
 quadrupole scalar flux is not. In fact, our results seem to hint at quadrupole scalar emission being as important  (or even larger, especially at low frequencies) in $k$-essence than in FJBD theories with the same conformal coupling $\alpha$, as long as the strong-coupling scale is in the MeV range that we can simulate.
{\it If} this feature survives for $\Lambda\sim \Lambda_{\rm DE}$, $k$-essence
theories may become as fine-tuned as FJBD theory (which will wash out any effect on large-scale structure observations).
Considering for instance the relativistic double-pulsar system $J0737-3039$~\cite{Kramer:2006nb,Noutsos:2020uip,Kramer:2021jcw}, in FJBD the absence of scalar 
quadrupole radiation constrains $|\alpha|\lesssim 4.4\times 10^{-2}$.\footnote{LIGO/Virgo bounds on deviations from GR at the quadrupole/dipole level produce weaker (if any) constraints on $\alpha$.}

\begin{acknowledgments}
\textit{Acknowledgments:} M.B, L.t.H, M.C, and E.B. acknowledge support from the European Union's H2020 ERC Consolidator Grant ``GRavity from Astrophysical to Microscopic Scales'' (Grant No.  GRAMS-815673) and the EU Horizon 2020 Research and Innovation Programme under the Marie Sklodowska-Curie Grant Agreement No. 101007855
C.P. acknowledges support from European Union FEDER funds, the Ministry of Science, Innovation and Universities and the Spanish Agencia Estatal de Investigación grant PID2019-110301GB-I00. MB  acknowledges the support of the PHAROS COST Action (CA16214). We acknowledge the use of CINECA  HPC resources thanks  to the agreement between SISSA and CINECA. 
\end{acknowledgments}

\appendix
\section{Appendix A: Covariant evolution equations}\label{fieldeqs}
All the equations presented in this section are in the Einstein frame and therefore, to avoid the clutter of notation, we remove the tilde and leave it understood that all quantities are always in the Einstein frame.
Notice, however, that the results of the convergence tests in the next section are converted back to the physical Jordan frame.

The equations of motion for the metric and scalar field can be obtained by varying the action~\eqref{action}
leading to
\begin{align}\label{fieldeqs1}
G_{\mu\nu}=8 \pi G\big( T^{\phi}_{\mu\nu} + T_{\mu\nu}\big)\;,\\\label{fieldeqs2}
\nabla_\mu\left[K'(X) \nabla^\mu \phi\right]=\frac{1}{2}\mathcal{A} T\;,
\end{align}
where $G_{\mu\nu}$ is the Einstein tensor, $\mathcal{A} \equiv -\Phi'(\phi)/[2\Phi(\phi)]$ and `prime' denotes a derivative with respect to the argument in parenthesis. The scalar field and matter energy-momentum tensors are defined as
\begin{eqnarray}
T^{\phi}_{\mu\nu} &=& K(X)g_{\mu\nu}-2K'(X) \de_\mu \phi \de_\nu \phi\; ,\\
T_{\mu\nu} &=& \frac{2}{\sqrt{-g}}\frac{\delta S_{m}}{\delta g_{\mu\nu} }\;,    \label{SET}
\end{eqnarray}
with $T=T_{\mu\nu}g^{\mu\nu}$.

For our purposes, the matter is described by using the stress-energy tensor for a perfect fluid, namely
\begin{equation}
T_{\mu\nu} = [\rho_0(1+\epsilon) + P]u_{\mu}u_{\nu} + P\,g_{\mu\nu}\,,
\end{equation}
where $\rho_0$ is the rest-mass density, $\epsilon$ its specific internal energy, $P$ is the pressure and  $u_{\mu}$ its four-velocity.

Finally, the general relativistic hydrodynamics equations are given by
\begin{eqnarray}
\nabla_{\mu}T^{\mu\nu} &=&\mathcal{A}\,T\,\nabla^\nu \phi  \,,
\label{fieldeqs4}
\\
\nabla_{\mu}(\rho_0 u^{\mu}) &=& \rho_0\mathcal{A}\,u^\mu \nabla_\mu\phi   \,. \label{fieldeqs3}
\end{eqnarray}
Note that these  are not conservation equations due to the coupling between the scalar field and matter.

We can now proceed to review the evolution equations for the metric, the scalar field and the matter, obtained by performing a 3+1 decomposition of the covariant equations.

\subsection{Metric}
The Einstein equations~\eqref{fieldeqs1} are written as an evolution system by using the covariant conformal Z4 (CCZ4) formulation, which extends the Einstein equations by introducing a four-vector $Z_{\mu}$ as follows
\begin{eqnarray}
&& R_{\mu\nu} + \nabla_{\mu} Z_{\nu} + \nabla_{\nu} Z_{\mu}   = 
8\pi \, \Big[ \left(T^{\phi}_{\mu\nu} + T_{\mu\nu}\right)  \\
&& - \frac{1}{2}g_{\mu\nu} \,\trT \Big]
+ \kappa_{z} \, \left(  n_{\mu} Z_{\nu} + n_{\nu} Z_{\mu} - g_{\mu\nu} n^{\sigma} Z_{\sigma} \right)\,, \nonumber
\label{Z4cov}
\end{eqnarray}
where  $\kappa_{z} > 0$ is a damping term enforcing the dynamical decay of the constraint violations associated with $Z_{\mu}$~\cite{Gundlach:2005eh}. 

As  customary, we split the spacetime tensors and equations into their space and time components by using  a $3+1$ decomposition. The line element is then given by
\begin{equation}
ds^2 = - N^2 \, dt^2 + \gamma_{ij} \bigl( dx^i + \beta^i dt \bigr) \bigl( dx^j + \beta^j dt \bigr), 
\label{3+1decom}  
\end{equation}
where $N$ is the lapse function, $\beta^{i}$ is the shift vector, and $\gamma_{ij}$ is the induced metric on each spatial foliation, denoted by $\Sigma_{t}$. In this foliation, we can define the normal to the hypersurfaces $\Sigma_{t}$ as $n_{\mu}=(-N,0)$ and the extrinsic curvature $K_{ij} \equiv  -\frac{1}{2}\mathcal{L}_{n}\gamma_{ij}$,  where $\mathcal{L}_{n}$ is the Lie derivative along  $n^{\mu}$.  By introducing a (spatial) conformal decomposition and some definitions, namely
\begin{eqnarray}
\bar{\gamma}_{ij} &=& \chi\,\gamma_{ij} ~~,~~
\bar{A}_{ij}      = \chi\left(K_{ij}-\frac{1}{3}\gamma_{ij} \trK \right),\\
\trKhat &=& \trK - 2\, \Theta ~~,~~
{\hat \Gamma}^i = {\bar \Gamma}^i + 2 Z^{i}/\chi,
\end{eqnarray}  
where $\bar\Gamma^i=-\partial_{j}\bar{\gamma}^{ji}$, $\trK=\gamma^{ij}K_{ij}$ and $\Theta \equiv - n_{\mu} Z^{\mu}$, one can obtain the final equations for the evolution fields $\{ \chi, \bar{\gamma}_{ij}, \trKhat, \bar{A}_{ij}, {\hat \Gamma}^i, \Theta  \}$. The explicit form of these equations is lengthy and can be
found in~\cite{Bezares:2017mzk}.  Note that the definitions above lead to new conformal  constraints, 
\begin{equation}
\bar{\gamma} = \det (\bar{\gamma}_{ij})= 1\,, \qquad \trAtilde  =\bar{\gamma}^{ij}\bar{A}_{ij}=0,
\end{equation}
which can also be enforced dynamically by including additional
damping terms proportional to
$\kappa_c >0$ in the evolution equations.  Finally, we supplement these equations with gauge conditions for the lapse and shift. We use the 1+log slicing condition~\cite{Bona:1994dr} and the Gamma-driver shift condition~\cite{Alcubierre:2002kk}, namely
\begin{eqnarray}
\partial_{t}N &=& \beta^{k}\partial_{k}N - N^{2}f(N)\trKhat \,, \\
\partial_{t}\beta^{i} &=& \beta^{k}\partial_{k}\beta^{i} + \eta\beta^{i} - N^{2}g(N){\hat \Gamma}^i \,,
\end{eqnarray}
being $f(N)$ and $g(N)$ arbitrary functions depending on the lapse, and $\eta$ a constant parameter. In our case, a successful choice is $f(N)=2/N$, $g(N) = 3/(4 N^2)$ and $\eta = 2$. 

\subsection{Scalar field}

The time evolution equation for the scalar field can also be obtained by performing a 3+1 decomposition on Eq.~\eqref{fieldeqs2}. Due to the genuinely non-linear character of this equation,
shocks might appear during the evolution, even if the initial data is smooth. This problem is also present in fluid dynamics, where it is solved by considering the weak (or integral) balance law form of the equations and then by using suitable High-Resolution-Shock-Capturing (HRSC) numerical methods to solve them. We follow here the same 
procedure, writing Eq.~\eqref{fieldeqs2} in balance law form
\begin{eqnarray}
\partial_{t}\vec{\mathcal{U}} + \partial_{i} F^{i}(\vec{\mathcal{U}}) = S(\vec{\mathcal{U}})\;,
\end{eqnarray}
where $\vec{\mathcal{U}}$ is a vector containing all the conserved or evolved fields, $F^{i}(\vec{\mathcal{U}})$ are the fluxes and $S(\vec{\mathcal{U}})$ the sources. Notice that  neither the fluxes nor the sources contain derivatives of the evolved fields.

By defining the following new evolved fields
\begin{eqnarray}
\phi_i&\equiv&\partial_{i}\phi~,\\
\Pi&\equiv&-n_\mu\partial^\mu\phi=-\frac{1}{N}(\partial_{t}\phi-\beta^{i}\phi_{i})~,
\end{eqnarray}
the evolution equations for the scalar field in terms of the 3+1 quantities can be written as
\begin{eqnarray}
\partial_{t}\phi   & = & \beta^k \phi_{k} -N\Pi~,\\
\partial_{t}\phi_i & = &-\partial_i(- \beta^k \phi_k +  N\Pi)~,\\
\partial_{t}(\sqrt{\gamma}\Psi) &=& - 
\partial_{k}[\sqrt{\gamma}(-\beta^k \Psi + N K'(X) \gamma^{kj}\phi_j)]\nonumber\\
&&+\frac{1}{2}N \sqrt{\gamma} \mathcal{A} T~,
\end{eqnarray}
where $\gamma = \det(\gamma_{ij})$ and $\Psi=K'(X)\Pi.$

Therefore, our conservative evolved fields are $\{\phi,\phi_{i},\Psi\}$, while  the primitive physical variables required to compute the fluxes and sources are given by $\{\phi, \phi_{i}, \Pi\}$. In order to calculate $\Pi$ from $\Psi$, the following non-linear equation
\begin{equation}
K'(X)\Pi - \Psi=0
\end{equation} 
 must be resolved numerically (for instance with a Newton-Raphson method) at each time-step.

\subsection{Matter}
The general relativistic hydrodynamics equations are  written in flux-conservative form as well, i.e.
\begin{widetext}
\begin{eqnarray}
\partial_{t}(\sqrt{\gamma}D) +\partial_{k}[\sqrt{\gamma}D(N v^k-\beta^k)] &=&N\mathcal{A} \sqrt{\gamma} D(-\Pi+v^k \phi_k) \,, \\
\partial_{t}(\sqrt{\gamma}\tau)
+\partial_{k}[\sqrt{\gamma}(-\beta^k \tau+ N[S^k-D v^k])] &=&
N \sqrt{\gamma} \mathcal{A}  \left[ \Pi T + \Pi D - v^k \phi_k D \right] +\sqrt{\gamma} \left[N S^{ij} K_{ij}-S^i\partial_{i}N\right] \,, \\
\partial_{t}(\sqrt{\gamma}S_i)
+\partial_{k}[\sqrt{\gamma}(-\beta^k S_i + N S^k_i)] &=& N \sqrt{\gamma} \mathcal{A} \left[ \phi_i T \right]  +\sqrt{\gamma} \left[N\Gamma^j_{ik}S^k_j+S_j\partial_{i}\beta^j-(\tau+D)\partial_{i}N\right] \,,
\end{eqnarray}
\end{widetext}
where $v^{i}$ is the fluid velocity measured by an  Eulerian observer
\begin{eqnarray}
v^{i} &=& \frac{u^{i}}{W} + \frac{\beta^{i}}{N},
\end{eqnarray}
with $W \equiv 1/\sqrt{1-\gamma^{ij}v_{i}v_{j}}$  the Lorentz factor. The evolved conserved variables in this case are the rest-mass density measured by an Eulerian observer ($D$), the energy density (with the exclusion of the mass density) ($\tau$) and the momentum density ($S_{i}$), which are defined in terms of the primitive field as follows:
\begin{eqnarray}
D &=& \rho W~, \\
\tau &=&  h W^2 -p - D~, \\
S_{i} &=&  h W^2 v_{i}~, \\
S_{ij} &=& \frac{1}{2} \left(v_i S_j + v_j S_i \right) + \gamma_{ij} P~, 
\end{eqnarray}
being $h=\rho_{0}(1+\epsilon) + P$ the enthalpy. Again, the relation between the evolved and the primitive fields is nonlinear and it needs to be solved numerically by using a Newton-Raphson method in order to calculate the fluxes and sources.

\section{Appendix B: Code validation}
The validity of our 3+1 evolution code evolving NSs in $k$-essence theory has been checked with two types of tests. First, we compared our simulations with solutions obtained by solving for static NSs in spherical symmetry. We also compared with simulations of perturbed stars, which were performed with an independent 1+1 code in spherical symmetry \cite{terHaar:2020xxb,Bezares:2021yek}. Here, we considered not only small oscillations, but also cases where stellar perturbations are large enough to induce the gravitational collapse of the star to a black hole. All these results are summarized briefly in the main text, and will not be repeated here. Second, we have performed standard convergence tests both for these isolated NS solutions, as well as for the binary NS coalescence simulations.

\begin{figure}
    \includegraphics[width=0.45\textwidth]{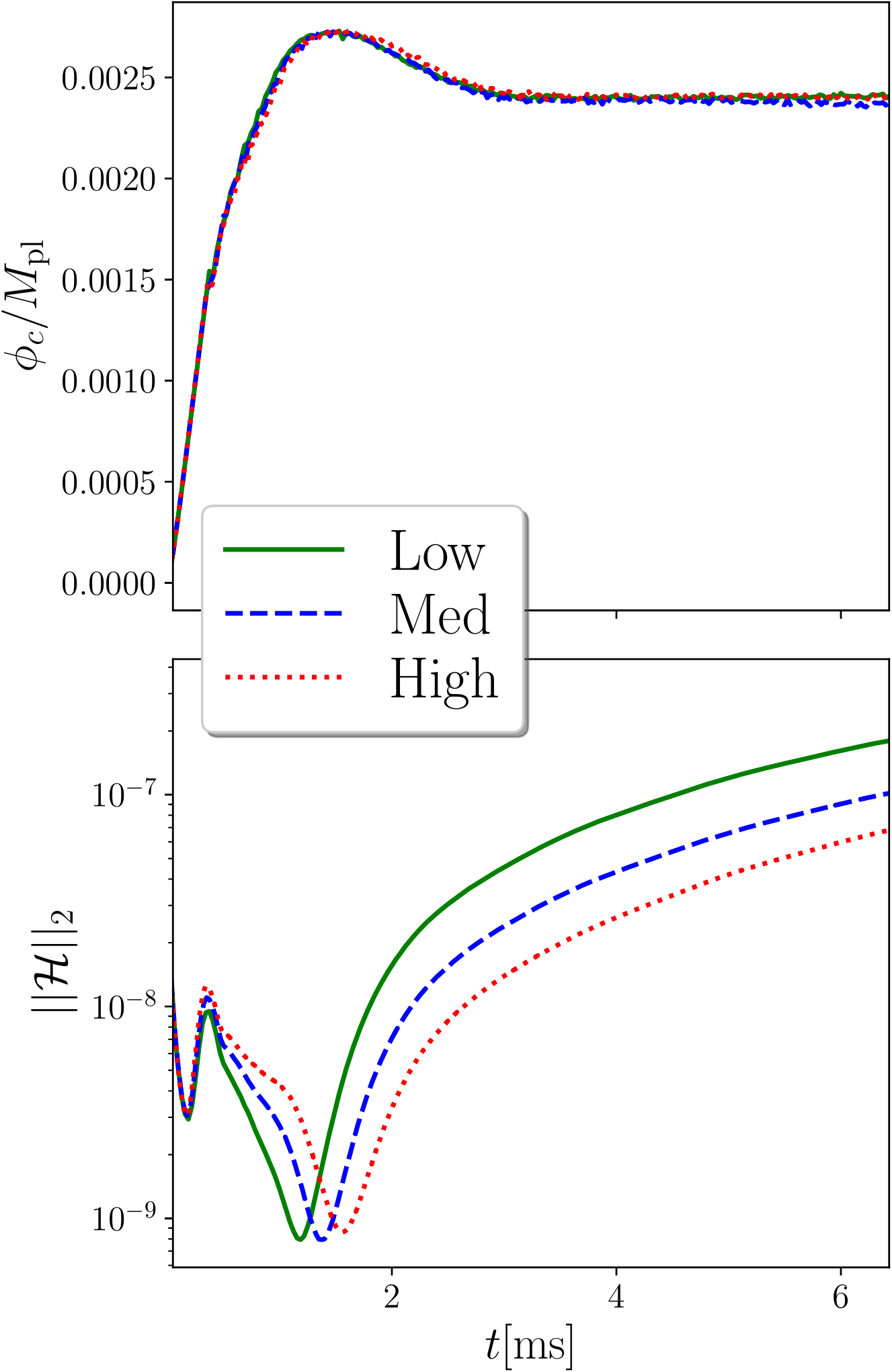}	
	\caption{{\it Isolated non-rotating NSs.} {\it Top}: Central value of the scalar field. The NS is initially a solution of Einstein equations, but it relaxes to the k-essence solution dynamically. We considered three different resolutions covering the star with $N=(80,100,126)$ points. {\it Bottom}:  L$_{2}$-norm of the Hamiltonian constraint for the three different resolutions above.
	After the transient initial relaxation, these solutions show between third and fourth order convergence. }
	\label{phihamisol}
\end{figure}

The first convergence test was performed for an isolated non-rotating NS with $\Lambda=4.0~\rm{MeV}$. We consider three different resolutions (i.e. low, medium and high) such that the star is covered roughly with $N=(66,82,103)$ points respectively. As described in the main text, our simulations begin from a NS solution in GR, which relaxes dynamically to  a $k$-essence solution. We display the value of the scalar field at the center of the star in the top panel of Fig.~\ref{phihamisol} for our three resolutions, showing that they all tend to the same value, which is very close to the one obtained with the 1+1 spherically symmetric code. Furthermore, the L2-norm of the Hamiltonian constraint for the three previous resolutions is displayed in the bottom panel of Fig.~\ref{phihamisol}, showing a convergence between third and fourth order.

\begin{figure}
    \includegraphics[width=0.45\textwidth]{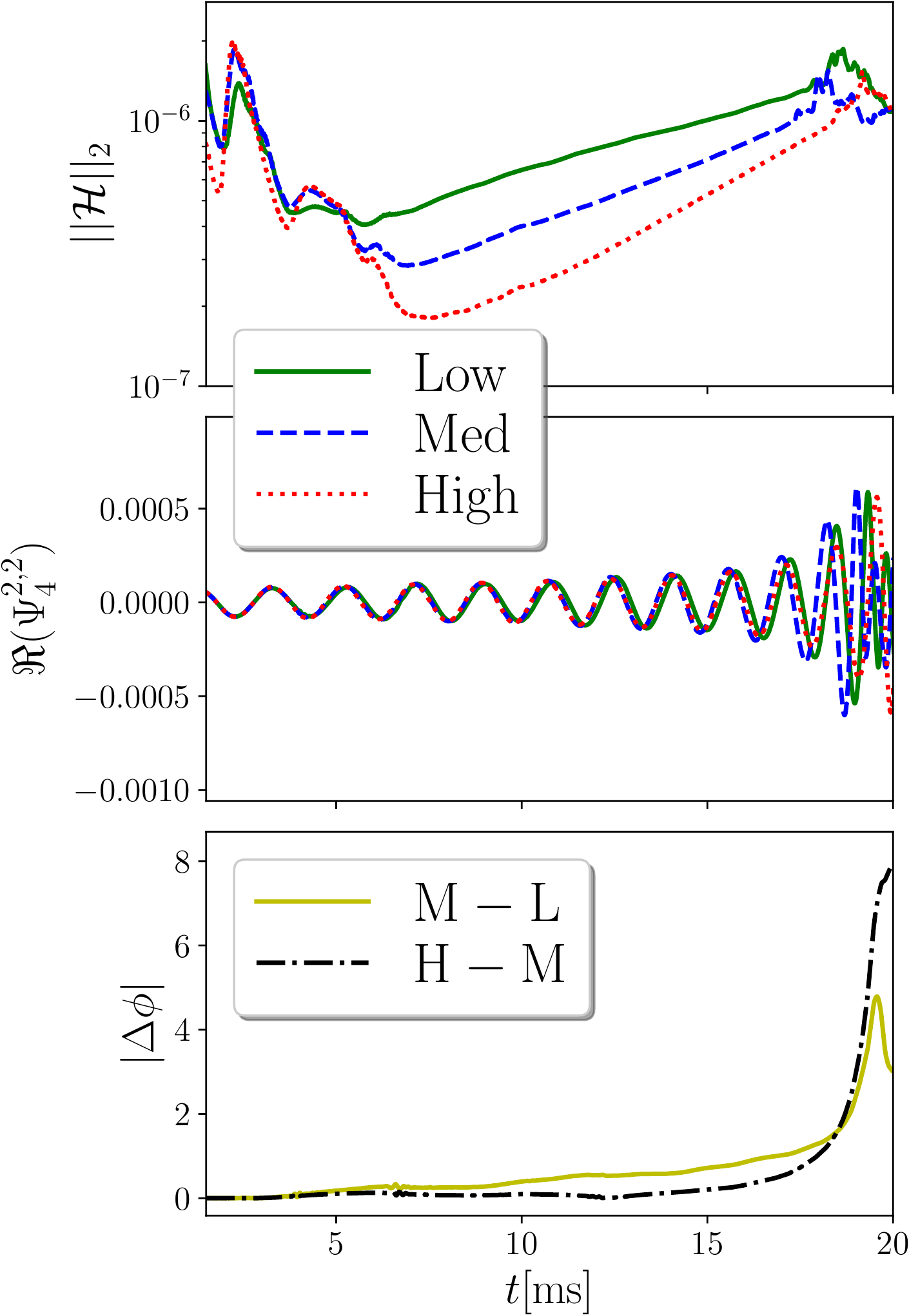}	
	\caption{{\it Binary NS simulations.} {\it Top}: The L2-norm of the Hamiltonian constraint for three resolutions $\Delta x=(366,292,232)~\rm{m}$. {\it Middle}: Main quadrupolar mode of the $\Psi_4$ for the three resolutions. {\it Bottom}: Difference between the phase of the different resolutions. The solutions converge at least at second order after the transient relaxation (from the GR solution to the $k$-essence one) and until the merger. }
	\label{conv_binary}
\end{figure}

The second convergence test was performed for one of our unequal binary NS simulations
with mass ratio $q=0.9$, using again  $\Lambda=4.0~\rm{MeV}$. We consider three resolutions $\Delta x=(366,292,232)~\rm{m}$, such that each star is covered roughly by $N=(80,100,126)$ points. In the top panel of Fig.~\ref{conv_binary} the L2-norm of the Hamiltonian constraint is displayed for these resolutions. The middle panel
displays the main quadrupolar mode $l=m=2$ of the GW signal. The differences between simulations decrease as the resolution increases. This is more evident from the bottom panel, where phase differences are displayed. Both the waveforms and the constraint converge, after the initial relaxation time and until almost the merger time, at least to second order. 

\bibliographystyle{apsrev4-1}
\bibliography{master}
\end{document}